# $B_1^+$ Homogenization in 7T MRI Using Mode-shaping with High Permittivity Materials

Yunchan Hwang†, Hansol Noh†, Minkyu Park, Jongho Lee, Namkyoo Park*

*Abstract*—Ultra high field (UHF) brain MRI has proved its value by providing enhanced SNR, contrast, and higher resolution derived from the higher magnetic field ($B_0$). Nonetheless, with the increased $B_0$ of UHF MRI, the transmit RF magnetic field ($B_1^+$) inhomogeneity also became one of the critical issues requiring attention. As the effective wavelength of RF becomes comparable or smaller than the dimension of the brain at $B_0$ larger than 7 Tesla, the increased $B_1^+$ inhomogeneity of UHF MRI results in poor SNR and uneven contrast. While parallel transmission techniques (PTx) and high permittivity material (HPM) structures for the mitigation of $B_1^+$ inhomogeneity have been suggested, the associated complexity in PTx and restricted volume of homogenization with HPM approach still remain as challenges. In this work, we address the $B_1^+$ inhomogeneity in the notion of mode-shaping. Treating a brain phantom as a dielectric potential-well resonator, we apply a phantom-conformal HPM potential-well in combination with a low-index potential barrier (air), to achieve the homogeneity of $B_1^+$ in the region of interest (ROI). Based on the electromagnetic simulations using a realistic brain model at 7T, we show that the proposed HPM structure reduces both the average deviation of $B_1^+$ in axial slices by 54% and peak SAR by 42%, respectively.

## I. INTRODUCTION

MRI has become a standard neuroimaging modality with its non-invasive nature and great flexibility in the domain of diagnostics and treatment planning. Since the advent of MRI in the 1980s, substantial efforts have been made to increase the magnitude of the main magnetic field ($B_0$), in order to achieve enhanced SNR, spatial resolution, and contrast [1-4]. The recent development of the ultra-high field (UHF, 7T or higher) MRI now allows the detection of subtle abnormalities associated with neurological diseases including multiple sclerosis, Alzheimer's disease, and brain tumors [5-8]. MR spectroscopy also benefits from the increased main field, providing improved spectral resolution compared to 3T MRI [9,10]. Meanwhile, the increase of main magnetic field $B_0$ also leads to critical penalties in the imaging, such as transmit RF magnetic field ($B_1^+$) inhomogeneity, which could result in poor SNR and uneven contrast [11-14]. Since the Larmor frequency scales with $B_0$, the effective wavelength of the $B_1^+$ field in the brain becomes smaller than the dimension of the brain at $B_0$ above 7T. Due to the phase evolution and interference of $B_1^+$ fields in the ROI, the $B_1^+$ inhomogeneity usually manifests in a fashion that the central region of the brain becoming brighter than the peripherals [15].

The most common attempt investigated for the mitigation of $B_1^+$ inhomogeneity includes parallel transmission techniques (PTx) and RF pulse modification. PTx utilizes RF control over amplitude, phase, and timing on each transmission element to achieve greater $B_1^+$ homogeneity, but at the expense of computational cost and hardware complexity [16,17]. $B_1^+$ inhomogeneity becomes particularly problematic in spin-echo based imaging such as $T_2$-weighted imaging, where magnetization refocusing is significantly affected by $B_1^+$ inhomogeneity [13,17-20]. Considering clinical and research significances of spin-echo sequences in $T_2$-weighted imaging and FLAIR imaging [12,21-23], intensive efforts have been made to improve $B_1^+$ homogeneity [24-28].

The use of high permittivity material (HPM) is an alternative approach that directly molds $B_1^+$ distribution and remains applicable in combination with spin-echo sequences and other homogenization techniques under a single- or multi-channel MRI environment [29,30]. High permittivity material such as titanate powder, when mixed with deionized water, generates extra magnetic fields in its vicinity [31,32]. HPMs in a polypropylene container usually called as "dielectric pads" [33] or HPMs with metallic structures a.k.a. "hybrid metasurface" generating even a greater extra magnetic field [34], thus have been extensively studied to improve SNR or $B_1^+$ field homogeneity when placed in the proximity of ROI [35-37]. While HPM pads have been adopted in numerous 7T applications including fMRI, diffusion imaging and structural imaging [38-41], $B_1^+$ enhancement and homogenization were limited to the vicinity of the structures, such as the cerebellum and temporal lobe, often resulting in the degradation of global $B_1^+$ homogeneity [17,42,43].

In this paper, we propose a "conformal, high permittivity" HPM, which supports robust boundary condition deriving global $B_1^+$ homogeneity in the whole brain region. Interpreting the HPM as a potential well which molds a boundary condition of a phantom, we show that HPM-permittivity value sufficiently higher than that of biological tissues preserves a robust boundary condition when evanescently-coupled to the phantom. To maximize $B_1^+$ homogenized ROI, a phantom-conformal boundary condition was established by using a cap-cylindrical HPM, consisting of a cylindrical HPM enclosing the head and a disk HPM placed in proximity of the apex of the head. Using a realistic head model of heterogeneous permittivity distribution, our approach confirms the reduction of $B_1^+$ magnitude deviation on average by 54% for all axial slices covering the region between the cerebellum and the top of the cerebrum. The max-to-minimum ratio of $B_1^+$ in an axial plane was also kept less than 2 throughout the cerebrum. Head average SAR and peak SAR were reduced by 32% and 42%, respectively.

†Yunchan Hwang and Hansol Noh contributed equally to this work
Yunchan Hwang, Hansol Noh, Minkyu Park and Namkyoo Park are with Photonic Systems Laboratory, Department of Electrical and Computer Engineering, Seoul National University, Seoul, Korea
Jongho Lee is with Laboratory for Imaging Science and Technology, Department of Electrical and Computer Engineering, Seoul National University, Seoul, Korea.
*Corresponding author: Namkyoo Park (nkpark@snu.ac.kr)

## II. PRINCIPLES

Instead of the conventional viewpoint treating the given problem of $B_1^+$ inhomogeneity in terms of electromagnetic wave transmission and interference, here we take the notion of the electromagnetic variational theorem [44], which fosters better intuitive interpretation. According to the theorem, an electromagnetic field concentrates in high permittivity region, which is often treated as an electromagnetic potential well. In this view, a typical $B_1^+$ inhomogeneity pattern, brighter in the central region and darker in the peripheral, is a mere manifestation of a field profile distribution (modal shape) in a high permittivity *multi-mode resonator* under external field excitations. Compensating for the $B_1^+$ inhomogeneity in this perspective hence corresponds to the spatial shaping of the mode, with the adjustment of the boundary condition and near-field couplings to the phantom, exercised by either a multi-channel RF coil or HPM. In the application of HPM, we emphasize that relatively higher permittivity materials providing strong field confinement are favored, in order to make the boundary condition robust against the presence of biological tissues.

## III. METHODS

EM simulation was performed using Transient solver of CST studio suite. We used a sphere (radius = 9 cm, $\varepsilon_r$ = 64 - modeling relative permittivity of grey matter at 300 MHz, and $\sigma$ = 0.2 S/m) as a phantom and MIDA, a detailed anatomical head and neck model, as a simulated brain, with corresponding physical property of biological tissue at 300 MHz [45,46]. In order to generate $B_1^+$, an ideal RF coil composed of a PEC shield and 8 PEC lines connected to current sources was driven in a $CP^+$ mode (Fig. 1 (a), shield radius = 18 cm, radial distance between the central axis and the current sources = 14.5 cm, height = 20 cm). Each PEC line (radius = 1 mm) was fed by two current sources, for their other ends were terminated with thin PEC disks of the same radius. Fig. 1(b) shows that the hard-sources driven coil achieves uniform $B_1^+$ field when unloaded, with a standard deviation of less than 4% in the loading region (radius = 9 cm, height = 20 cm). The current in each PEC line remains unchanged upon the loading of the phantom or HPMs, which allowed us to experiment with various HPM structures without tuning before $B_1^+$ inhomogeneity evaluation. In the phantom simulation, we focused on resolving $B_1^+$ inhomogeneity in the upper half of the phantom, whose size is comparable to that of the brain (Fig. 1(c)).

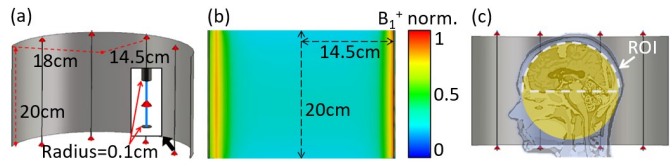

**Figure 1. Idealized RF coil driven by hard sources.**
a) Schematic of the RF coil in sagittal view. b) Sagittal $B_1^+$ color map of the unloaded RF coil. c) ROI of the sphere phantom with overlapped simulated brain voxel sagittal image. The upper half of the phantom's size is comparable to that of the brain.

## IV. RESULTS

### A. Phantom

Figure 2 illustrates the effect of a boundary condition on the dielectric phantom, exercised by the application of HPMs at different representative permittivity values. At first, the spherical phantom and a cylindrical HPM (inner radius = 10.5 cm, outer radius = 12.2 cm, height = 20 cm, no conductivity) were assumed for simulation. Considering the modal field distribution of the HPM in the absence of phantom, the radial distance between the HPM and the phantom was set to 1.5 cm, to achieve sufficient but not too strong couplings in between.

Figure 2 (a) and (b) show the case of HPM with comparable permittivity ($\varepsilon^{HPM}$ = 64) to the phantom ($\varepsilon_r$ = 64), failing to retain its $B_1^+$ field and boundary condition upon loading of the phantom. In contrast, when $\varepsilon^{HPM}$ was set sufficiently higher than the phantom's, the $B_1^+$ field inside the HPM remained essentially unchanged upon phantom insertion, as shown in Fig. 2 (c)-(h). Scanning through optimal HPM permittivity, we then selected $\varepsilon^{HPM}$ = 200, which provided a well-confined, robust $B_1^+$ field boundary condition in the HPM (Fig. 2(c)) as well as moderate $B_1^+$ excitation in the phantom (Fig. 2(d)). It is noted that too high permittivity HPM ($\varepsilon^{HPM}$ = 230, Fig. 2 (e), (f)) was excluded due to its very weak $B_1^+$ coupling to the phantom.

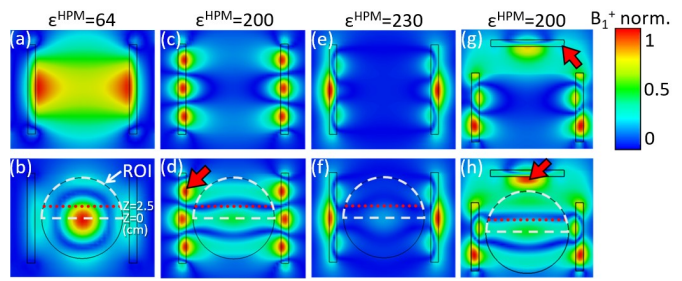

**Figure 2. Effect of boundary condition adjustment on $B_1^+$ mode of the idealized dielectric phantom.**
The top panels and the bottom panels show sagittal $B_1^+$ field profiles of the HPMs without and with the loading of the phantom, respectively. a,b) For HPM permittivity $\varepsilon^{HPM}$ similar to the phantom's ($\varepsilon_r$ = 64), the phantom-loaded HPM boundary condition was significantly disturbed from that of phantom-unloaded case. c~h) For $\varepsilon^{HPM}$ > $\varepsilon_r$, the HPM boundary condition was less susceptible to the presence of the phantom. c,d) With $\varepsilon^{HPM}$ = 200, the HPM was optimized for enhanced $B_1^+$ in the phantom's peripheral. e,f) With $\varepsilon^{HPM}$ >> $\varepsilon_r$, HPM-phantom coupling became too weak to excite $B_1^+$ in the phantom. g,h) "Conformal" boundary condition established by the cap-HPM and cylindrical-HPM ($\varepsilon^{HPM}$ = 200). Enhanced axial $B_1^+$ homogeneity was obtained when compared to the cylindrical-HPM.

The "cylindrical" HPM of $\varepsilon^{HPM} = 200$ produced relatively homogeneous $B_1^+$ field in *transverse direction*, as shown in Fig. 2 (d) with reduction in $B_1^+$ max-to-minimum ratio (M/m) at $z = 0$ from 10 (Fig. 2 (b)) to 1.5. (Fig. 2 (d)). However, $B_1^+$ homogeneity was partially degraded around curved nodal surfaces ($z = 2 \sim 3$ cm) of $B_1^+$. In the axial plane at $z = 2.5$ cm, marked with a red-dotted line in the bottom panel of Fig. 2, no significant reduction in M/m was observed by application of the cylindrical HPM (6.4 and 6.3 in Fig. 2 (b) and (d), respectively). Focusing on the inhomogeneity in the upper half-sphere of the phantom, the $B_1^+$ inhomogeneity could be further mitigated by applying the HPM boundary condition conformal to the upper half-sphere. In order to achieve a "conformal" boundary condition for the ROI, we placed an HPM "cap" disk (marked with a red arrow in Fig. 2 (g)) above a shortened cylinder, shifting the position of field confinement: from the top region of the un-shortened cylindrical HPM (arrow in Fig. 2 (d)), to the center of the "cap" HPM (arrow in Fig. 2 (h)). Fig. 2 (g) shows $B_1^+$ field distribution of the cap-cylindrical HPM configuration (HPM disk $\varepsilon^{HPM} = 200$, radius = 9 cm, thickness = 1.55 cm, positioned 5.7 cm above the cylindrical HPM of 15 cm height). With the $B_1^+$ field excitation from the cap-cylindrical HPM to the upper half-sphere of the phantom, the nodal plane in the ROI was flattened with M/m = 2.0 at $z = 2.5$ cm (Fig. 2(h)), achieving transversely homogeneous $B_1^+$ field throughout the upper half-sphere.

Figure 3 compares the results of different HPM structure applications to the phantom. Sagittal $B_1^+$ color maps, normalized in each phantom, are shown in Fig. 3 (a)-(c) for the cases of phantom only (no HPM), phantom with cylindrical HPM, and phantom with cap-cylindrical HPM, respectively. The boundary conditions imposed by the HPM structures provided a relatively homogenous $B_1^+$ field in the phantom's axial planes, compared to the case of phantom only.

For the quantification of $B_1^+$ inhomogeneity in axial planes - widely used orientation in the clinical environment - we used two figures-of-merit (FOMs) in the ROI: max-to-minimum ratio (M/m) and coefficient of variation (CV, $\sigma/\mu$). Only the upper half of the phantom ($z = 0$ to 9 cm, with axial slices of 1-mm thick, 2.5-mm inter-slice distance) was considered in the FOM evaluation. Fig. 3 (d) and (e) show CV and M/m respectively along the superior direction. With the application of the cylindrical HPM, overall improvement in CV and M/m was achieved throughout the ROI, except the region around the curved nodal plane, near $z = 2.5$ cm. Best CV and M/m was obtained with the conformal, cap-cylindrical HPM as expected. In the ROI of the upper half-sphere, a substantial reduction in CV and M/m was observed, on average, by 42% and 45% respectively. Fig. 3 (f) also shows axial $B_1^+$ color maps that were normalized with respect to the average $B_1^+$ of each slice. With the flattening of the nodal plane through the progressive adjustment in boundary conditions, from spherical (case of no HPM, Fig. 3 (a)) to curved (case of cylindrical HPM, Fig. 3 (b)), then to flattened (case of cap-cylindrical HPM, Fig. 3 (c)) ones, $B_1^+$ field homogenization was sequentially achieved at $z = 2$ cm and 4 cm.

### B. Brain model simulation

Having verified the concept of $B_1^+$ homogenization under the notion of conformal mode-shaping with the phantom, here we examine the applicability of the same idea using the simulated brain, whose permittivity distribution is vastly heterogeneous in nature. For $B_1^+$ homogenization in the brain, we tested the cap-cylindrical HPM structure only, which proved the best performance in the phantom. Considering the size of the brain, the dimensions of HPMs were slightly adjusted in the optimization process. The disk (radius = 9 cm, thickness = 1.55 cm) was placed 9.1 cm above the cylinder (inner radius = 12.2 cm, outer radius = 13.5 cm and height = 17.8 cm). The cap-cylindrical HPM's permittivity was kept at $\varepsilon^{HPM}$ = 200 with loss tangent of 0.07, the average value of unpressured- and unsaturated- barium titanate mixture [47].

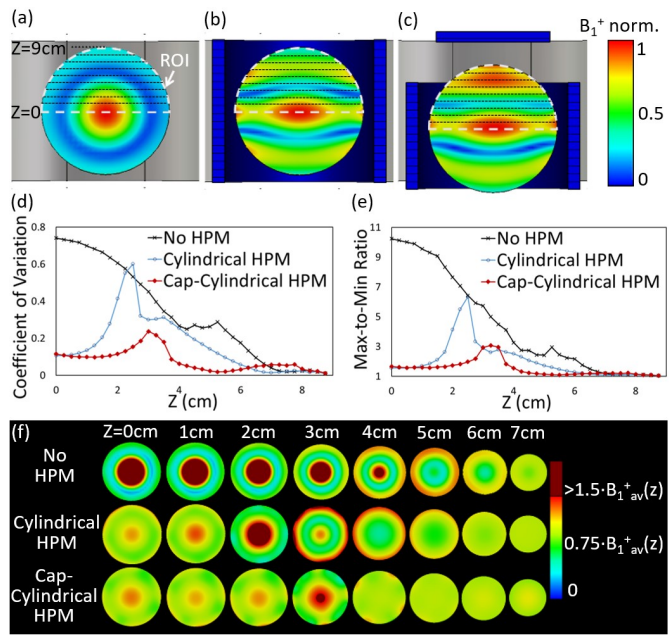

**Figure 3. Phantom-conformal mode-shaping with cap-cylindrical HPMs for $B_1^+$ inhomogeneity mitigation.**
Sagittal $B_1^+$ profiles of the phantom in the cases of a) no HPM, b) cylindrical HPM and c) cap-cylindrical HPM. d,e) Coefficient of Variation and M/m in axial slices along superior direction, at different $z \geq 0$. f) Representative axial $B_1^+$ images along superior direction, at different $z \geq 0$. $B_1^+$ was normalized in each axial slice.

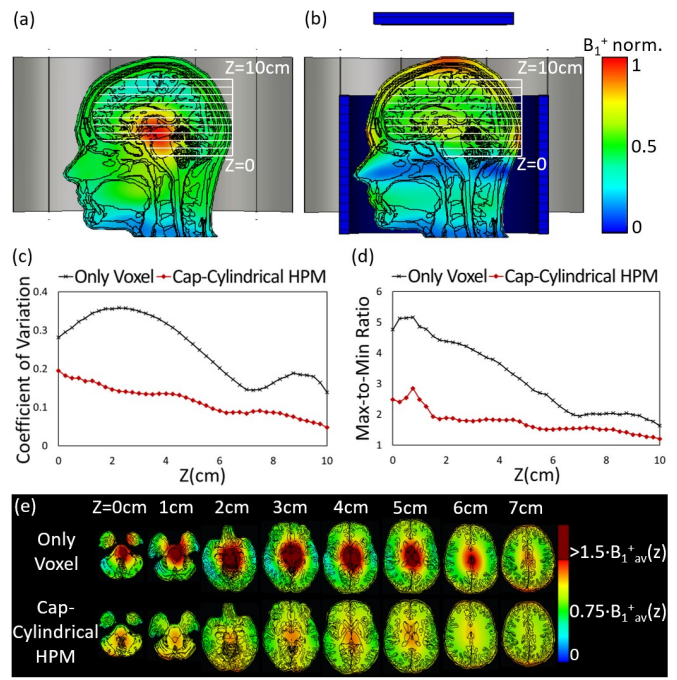

**Figure 4. $B_1^+$ inhomogeneity mitigation by mode-shaping on the voxel using HPMs.**
Sagittal $B_1^+$ profile of a) the voxel and b) the voxel with the cap-cylindrical HPM structure ($\varepsilon^{HPM} = 200$, $\tan\delta = 0.07$) c,d) Coefficient of Variation and M/m in axial slices along superior direction. e) Representative $B_1^+$ images along superior direction. $B_1^+$ was normalized in each axial slice.

Figure 4 shows the effect of cap-cylindrical HPM on $B_1^+$ homogeneity in the simulated brain voxel. $B_1^+$ color maps in Fig. 4 (a),(b), and (e) were normalized in the same manner as in Fig. 3 for the visualization of $B_1^+$ inhomogeneity. Fig. 4 (a) and (b) confirm the idea of boundary condition fixing with the high permittivity ($\varepsilon^{HPM}$ = 200) material; the overall feature of the modal profile observed in the spherical phantom (Fig. 3 (c)) was preserved, not significantly perturbed by the heterogeneous permittivity distribution of the brain. Fig. 4 (c) and (d) show CV and M/m respectively along the superior direction, where only brain tissues were considered in the evaluation. The cap-cylindrical HPM consistently provided improved $B_1^+$ homogeneity at every axial slice (1-mm thick, 2.5-mm inter-slice gap), between $z$ = 0 and 10 cm. The CV and M/m were reduced on average by 54% and 41% compared to the only-voxel case. Worth to mention, the FOM of M/m was suppressed well below less than 2 over the region between $z$ = 1.5 cm (middle of the cerebellum) and 10 cm (top of the cerebrum). Fig. 4 (e) shows eight pairs of axial $B_1^+$ color maps along the superior direction. The bright-center dark-peripheral patterns of the brain (upper panels) were suppressed to a great extent in the cerebellum, deep grey matter, and cerebrum altogether, with the use of the cap-cylindrical HPM.

## C. SAR analysis

The issue of SAR is also a critical feature that needs to be addressed in the application of HPM. To ensure that the cap-cylindrical HPM does not introduce significant RF heating, we compared SAR (10 g) between the only-voxel brain model case and the cap-cylindrical HPM case (Fig. 5). SAR was calculated employing the same procedure in Vaidya et al. [39]; average $B_1^+$ in the ROI was scaled to flip magnetization in 2 ms for both 90°, and 180° pulses and the square of the $B_1^+$ scaling was used to scale SAR. Assuming a Turbo spin-echo (TSE) sequence (TR=6000 ms, ETL=16), the cap-cylindrical HPM structure reduced the head average SAR by 32% (0.22 W/kg to 0.15 W/kg) and the peak SAR by 42% (0.79 W/kg to 0.46 W/kg), as illustrated in Fig. 5 (a) and (b). It is noted that the SAR reduction from the HPM application was most prominent at the center of the brain, and high SAR region was shifted closer to the skin (Fig. 5(c), $z$ = 3~6 cm).

## D. Robustness of the mode-shaping

In the clinical environment, displacement of the head could affect the $B_1^+$ homogeneity. The robustness of proposed $B_1^+$ homogenization with mode shaping was tested against three types of neck *rotation* and the variation of head size: neck flexion, extension, lateral flexion, and ±20% in volume (Fig. 6 (a)). Rotation angles were set 5° and 10° in each direction, and rotation axes crossed the first cervical vertebra. Fig. 6 (b) compares the average reduction of CV and M/m in six head displacements and two size variations to those of the neutral position. No significant change for the $B_1^+$ homogenization capability of the cap-cylindrical HPM was observed, with less than 3.5%p standard deviation in the average reduction of CV and M/m for all rotations and size variations. For all cases, CV reduction was higher than 47% (+20% larger head size), and M/m reduction was higher than 36% (10° flexion). We note that low permittivity spacer, such as polypropylene, can be further used to mitigate the shift and movement of the head.

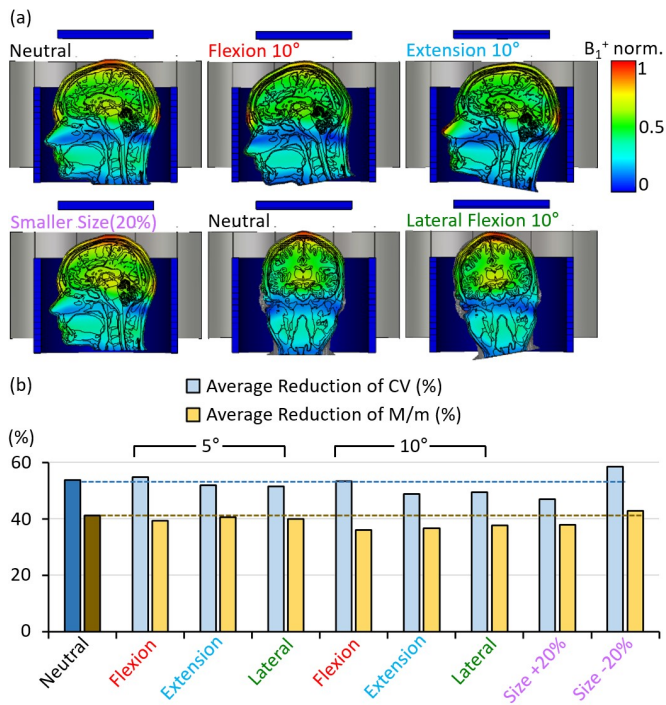

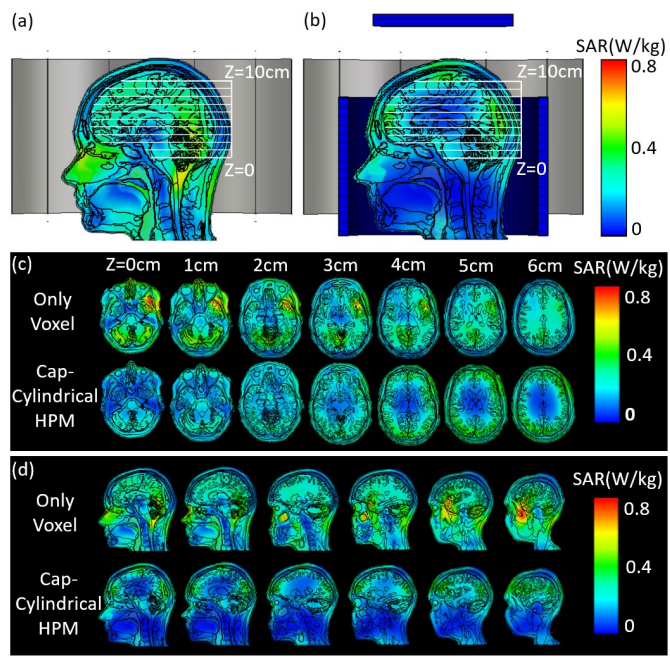

**Figure 5. Comparison of SAR in the voxel without and with the application of cap-cylindrical HPM.**
Sagittal SAR (10 g) profile corresponding to the TSE sequence for a) the voxel and b) the voxel with the cap-cylindrical HPM structure. c, d) Representative SAR images at different axial and sagittal positions.

**Figure 6. Robustness of $B_1^+$ homogenization.**
a) Robustness of the cap-cylindrical HPM's $B_1^+$ homogenization was tested against three types of rotations (neck flexion - front bending, extension - backward bending, and lateral flexion) and size variations. Rotation angles were set 5° and 10° and volume scaling factors were set +20% and -20%. b) Graph shows average reduction of inhomogeneity (CV and M/m) with the application of the cap-cylindrical HPM for each case. For all conditions, changes in the $B_1^+$ inhomogeneity reduction from the neutral position were less than 7% for both CV and M/m.

## V. Discussion

In this work, HPM was employed to *homogenize* $B_1^+$ in the whole brain, not to *enhance* $B_1^+$ in a local region in the vicinity of HPM as "dielectric pads" do. While cylindrical HPM in Fig. 2 (d) successfully mitigated axial $B_1^+$ inhomogeneity, the observed inhomogeneity along the superior direction, disconnected by the nodal $B_1^+$ planes, was found not ideal for brain MRI. To achieve a conformal boundary condition and $B_1^+$ homogeneity over the entire brain, a shortening of the cylindrical HPM was required with the addition of the cap HPM. We further note that the z-directional height of the homogenization region can be further extended by modifying the resonance mode in the cylindrical HPM: for example, by slightly lowering the effective index of HPM with the reduction of its thickness, while increasing its height.

Robustness of mode-shaping $B_1^+$ homogenization against head movement and size variation (Fig. 6) shows the feasibility of the cap-cylindrical HPM in real practice. This robustness also justifies our proposal of using relatively higher permittivity HPM, in controlled distance from the brain: which provides a well-confined, robust $B_1^+$ field boundary condition in the HPM (Fig. 2(h)) as well as sufficient $B_1^+$ excitation in the brain.

While barium titanate mixture well supports the assumed large permittivity of HPM (ε = 200) without a significant increase of conductivity from saturation, the structural integrity of the mixture can be problematic in clinical practice. Barium titanate mixture also can be deformed by stress and desiccated over time [47,48]. Instead, metamaterials providing equivalent values of effective permittivity could compose an attractive candidate with an inexpensive polyethylene casing, removing the risk of desiccation or leakage of toxic barium titanate mixture. With the long Larmor-wavelength (1 m) at $B_0 = 7T$, the design and fabrication of high permittivity metamaterial structure will not raise practical problems.

## VI. Conclusion

While HPM pads have been extensively studied for $B_1^+$ homogenization in UHF MRI, the target region was limited to the vicinity of the pads, often degrading universal homogenization over the extended ROI. In this work, we demonstrate that axial $B_1^+$ homogeneity can be improved in the whole brain ROI, by using a conformal HPM. Specifically, we established a robust boundary condition ideal for the $B_1^+$ homogenization in the ROI, by employing conformal HPMs of relatively high permittivity, placed in proximity for evanescent couplings to the phantom. The $B_1^+$ homogenization using the brain-conformal cap-cylindrical HPM structure remained valid for a realistic head model of heterogeneous permittivity distribution, providing M/m value of less than 2 throughout the cerebrum. Reduction in the head average SAR by 32% and the peak SAR by 42% was also realized. Since the $B_1^+$ distribution itself is homogenized without small tip angle assumption and a complex parallel transmission system, the proposed structure is expected to be most useful in spin-echo based imaging in UHF MRI, such as $T_2$-weighted imaging and diffusion-weighted imaging in the whole brain. Especially considering that TSE is deemed as a "workhorse" in the clinical environment, and noting the crucial role of $T_2$-weighted imaging, the concept of formulating a conformal boundary condition via an HPM structure may foster the clinical application of UHF MRI.